\documentclass[aps,prb,twocolumn,superscriptaddress]{revtex4}
\usepackage{ae}
\usepackage[T1]{fontenc}
\usepackage[ansinew]{inputenc}
\usepackage{amsmath}
\usepackage{amssymb}
\usepackage{graphicx}
\usepackage{color}
\usepackage[colorlinks]{hyperref}
\usepackage{epstopdf}

\begin{document}


\title{Magnetoconductivity of a metal with closed Fermi surface reconstructed by a biaxial density wave}

\author{A.M. Kadigrobov* }
\affiliation{Theoretische Physik III, Ruhr-Universitaet Bochum, D-44801 Bochum, Germany}
\affiliation{Department of Physics, Faculty of Science, University of Zagreb, Croatia}
\author{B. Keran}
\affiliation{Department of Physics, Faculty of Science, University of Zagreb, Croatia}
\author{D. Radi\'{c}$^\dagger$}
\affiliation{Department of Physics, Faculty of Science, University of Zagreb, Croatia}
\begin{abstract}

We investigate quantum dynamics and kinetics of a 2D conductor with closed Fermi surface reconstructed by a biaxial density wave, in which electrons move  along a two-dimensional periodic net of semiclassical trajectories coupled  by the magnetic breakdown tunnelling under a strong magnetic field.  We derive a quasi-particle dispersion law and magnetoconductivity tensor.
The quasi-particle spectrum is found to be the alternating series of  two-dimensional magnetic energy bands with gaps between them. The longitudinal magnetoconductivity shows giant oscillations with change of magnetic field, while the Hall coefficient changes sign and is absent in a wide range of the magnetic fields in between. Preliminary estimations show that the suggested magnetoconductivity mechanism may be the origin of such behaviour of the Hall coefficient vs. magnetic field, as observed in experiments in materials with analogous topology of the Fermi surface, such as the high-$T_c$ superconducting cuprates.
\end{abstract}

\maketitle

\section{Introduction \label{introduction}}

Systems with charge carriers interacting with phonons or other excitations in one- and  two-dimensional (1D and 2D) geometry are often found to be unstable with respect to spontaneous arising of a periodic modulation of charge or spin density which is usually called the density wave (DW) \cite{Gruner}. 
Such periodic ordering is stabilized whenever the decrease of the electronic band energy, which occurs due to the reconstruction 
of the Fermi surface (FS), overcomes the increase of the crystal energy caused by the crystal lattice modulation or electron-electron interactions. In 1D conductors such modulation opens a gap in an electron spectrum at the Fermi energy, thus decreasing the electron band energy, constituting the well-known Peierls instability \cite{Peierls}.
The DW instability is  most commonly encountered in the 2D conductors with highly anisotropic open FS, such that entire or great part of its contour can be mapped onto each other by a single wave vector. It is known as the nesting mechanism of the DW  stabilization \citep{Gruner,Pouget,Thorne}. 
The DWs have been also observed in conductors with closed and convex FSs, such as the high-$T_c$ superconducting cuprates \cite{DWcuprates}, or intercalated graphite compounds \cite{CaC6}, the DW instability of which can not be explained within the above-mentioned nesting mechanism. A possible explanation of this phenomenon was recently suggested in our papers \cite{DWPRB1,DWPRB2,DWPRB3}. This mechanism is based on the topological reconstruction of the initially closed FS into an open one, the latter being composed of the initial FSs which slightly overlap ("nearly touch") one another with (pseudo)gap opening around the "touching" region. 

Of particular interest are not only the DW properties,  but also  
a response of such system to external fields, temperature  or currents. Conductance and its dependence  on magnetic field are of great  importance both for fundamental science and  applications. Such phenomena  as the de Haas - van Alphen, or Shubnikov - de Haas oscillations, as well as the Hall effect, provide a powerful tool for the investigations  of  new characteristics of the material imposed by the DW such as, for example, the quasi-particle FS reconstructed by the DW. Change of sign of the Hall coefficient \cite{HallChange} and  its  zero-value in a wide interval of magnetic field \cite{HallAbsence}, observed in experiments on high-$T_c$ superconducting cuprates, are particularly intriguing.

 In this paper we investigate dynamics and kinetics of quasi-particles in a 2D conductor, with a DW ordering, under a strong magnetic field perpendicular to the sample. We consider the case of biaxial DW (so-called "chequerboard pattern"), characterized with two perpendicular wave vectors of the same size, which brings the initial closed FSs into a 2D net of FSs very slightly overlapping  ("nearly touching") one another. By lifting the energy degeneracy and opening the gap around the points of contact,  the FS is transformed into a periodic set of closed diamond-shaped pockets, which are close to each other at the edges of the new Brillouin zones in the extended zone scheme \cite{Comment} (see Fig.\ref{FigOrbitNet}).\ 
 
Under a strong magnetic field, quasi-particles move along the semiclassical trajectories undergoing magnetic breakdown (MB) in the small gapped area around the touching points (MB region shown shaded/dotted in Fig.\ref{FigOrbitNet}) with the MB probability \cite{MBpobabTouch1,MBpobabTouch2,Vorontsov,Fortin}
\begin{eqnarray}
|t|^2\approx 1- \exp{\left(-\frac{\Delta^2}{(\hbar \omega_H)^{4/3}\varepsilon_F^{2/3}}\right)}
\label{MBprobab}
\end{eqnarray}
to scatter and continue motion along the diamond-shaped orbit, and probability $|r|^2=1-|t|^2$ to tunnel through the semiclassically forbidden regions and move along the circular orbit. There $\omega_H =eH/m$ is the cyclotron frequency, $H$ is the magnetic field, $e$ is the electron charge (absolute value), $m$ is the electron effective mass, $\varepsilon_F$ is the Fermi energy and $\Delta$ is the gap in the quasi-particle energy spectrum.
This equation for the MB probability differs from the well-known Blount formula \cite{Blount,KS,Slutskin} $|t|^2=1-\exp{\left(-\Delta^2/(\hbar \omega_H |v_x^{(0)}v_y^{(0)}|)\right)}$ 
where $v_x^{(0)}$ and $v_y^{(0)}$ are the velocity projections  at the magnetic breakdown point between trajectories crossing each other before opening of the gap. This difference appears because one of the velocity projections in the Blount probability would be equal to zero if the MB takes place between the trajectories "touching" each other as in the present case.

We calculate spectrum of quasi-particles moving in such a 2D MB net under magnetic field \cite{2Dexperiment} satisfying the condition that $\Phi/\Phi_0$ is a rational number, where $\Phi$ is the magnetic flux threading the unit cell and $\Phi_0=h/e$ is the magnetic flux quantum ($h$ is the Planck constant). This condition, appearing in our problem, is analogous to the one from the well-known works of Zak \cite{Zak} and Hofstadter \cite{Hofstadter}, coming from the requirement that the wave function must be uniquely defined, i.e. from the required commensurability of the new periodicity imposed by the magnetic field with the periodicity of the underlying lattice imposed by the crystal potential or the DW.
Our analytical calculations show that the quasi-particle spectrum in such system is a serial of alternating 2D energy bands with gaps between them.

Using this spectrum we calculate the 2D magnetoconductivity tensor $\sigma_{ij}$, $i,j \in \{ x,y \}$ and find that the diagonal conductivities $\sigma_{xx}$ and $\sigma_{yy}$ perform giant oscillations with a change of magnetic field  in the whole interval of $H$ in which the MB is distinct, i.e. $|t||r|\sim 1 $. It is particularly remarkable that $\sigma_{xy}=0$ and hence the Hall coefficient $R_H=0$ everywhere inside this wide interval of magnetic field.

We also show that the conventional  magnetoconductivity tensor  for closed orbits \cite{LAK} restores in the limits of relatively weak, i.e. $|r(H)| \ll 1$,  and relatively strong  magnetic fields, i.e. $|t(H)| \ll 1$, in which broadening of the Landau levels by the MB is much smaller than the energy uncertainty caused by the quasi-particle - impurity scattering characterized by the scattering frequency $\nu_0$. The Hall coefficient has opposite signs in these two limits (and vanishes between them).

The paper is organized as follows: In Section \ref{SecDynamics} we 
consider dynamics of quasi-particles moving in the 2D periodic net of semiclassical trajectories coupled by the MB, finding the quasi-particle wave functions and dispersion law in the momentum representation. In Section \ref{SecKinetics} we calculate the magnetoconductivity tensor. Section \ref{SecConclusion} contains the concluding remarks. In Appendix we show mathematical details related to the calculation of the above-mentioned quantities.

\section{Dynamics of quasi-particles in a two-dimensional net of orbits  under magnetic breakdown conditions \label{SecDynamics}}

Recently, we have shown that a homogeneous state of electrons in 2D conductors with closed FSs may be unstable with respect to spontaneous arising of a uniaxial DW in the system \cite{DWPRB1,DWPRB2,DWPRB3}.
%
\begin{figure}
\centerline{\includegraphics[width=8.0cm]{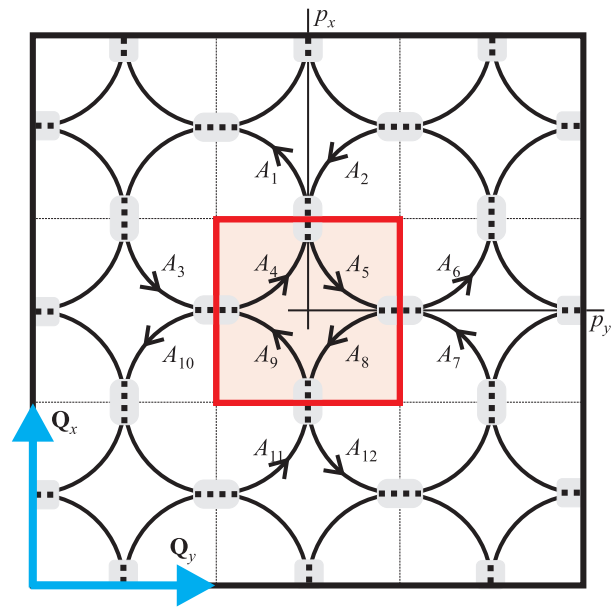}}
\caption {The Fermi surface (extended zone scheme) ofinvestigation of ) a 2D conductor reconstructed by the biaxial density wave (characterized by wave vectors $\textbf{Q}_x$ and $\textbf{Q}_y$ in $x-$ and $y-$direction). Initially circular contours (electron-like) are transformed into a set of diamond-like contours (hole-like) separated by gaps (shaded gray) in  small areas at the Brillouin zone boundary (red square). Under the magnetic field perpendicular to the plane, quasi-particles move in the momentum space $(p_x,p_y)$ along the semiclassical trajectories (arrows) undergoing the MB scattering at the gapped areas (dotted paths). $A_1 - A_{12}$ are constants related to the semiclassical wave functions on the corresponding trajectories in the 2D MB net. .}
\label{FigOrbitNet}
\end{figure}
%
In this work we consider a 2D conductor, with initial quasi-particle spectrum $\varepsilon(p_x,p_y)$ and momentum $\mathbf{p}=(p_x,p_y)$, and closed FS determined by $\varepsilon(p_x,p_y)=\varepsilon_F$, in which the  structural  instability is caused by a biaxial DW characterized by two perpendicular wave vectors $\textbf{Q}_x$ and $\textbf{Q}_y$ in $x-$ and $y-$direction respectively. We consider the case  which results in a reconstruction of the initial closed FSs  into a 2D periodic set of (new) FSs which are close to each other at the edges of the (new) Brillouin zones in the extended zone scheme as shown in Fig.\ref{FigOrbitNet}.
Under a strong magnetic field $\varepsilon_F/\hbar \gg \omega_H \gg \nu_0$,  the quasi-particle moves in the momentum space along the semiclassical trajectories at energy $\varepsilon(p_x, p_y) =\varepsilon$  between the MB regions (shaded in Fig.\ref{FigOrbitNet}), undergoing the MB scatterings at them\cite{comment1}.\

The semiclassical motion of the quasi-particle in the momentum space along  the semiclassical trajectories between the MB points is described by the Lifshitz-Onsager Hamiltonian \cite{Lifschitz,Onsager}. Choosing  the Landau gauge for the vector potential, $\mathbf{A}= (-H y,0,0)$, one writes the Schr\"{o}dinger  equation in the momentum representation for quasi-particles moving between the MB points
%
\begin{eqnarray}
\varepsilon_{\alpha}\left( P_x +i b^2_H \frac{d }{d p_y}, p_y \right) G_\alpha(p_y)=\varepsilon G_\alpha(p_y),
\label{LOSchrod}
\end{eqnarray}
%
where $\varepsilon_\alpha(p_x,p_y)$ is the initial  quasi-particle dispersion law which is shifted to the position corresponding to the trajectory $\alpha$ (the index $\alpha$ denotes the quasi-particle trajectory between two neighbouring MB points - see Fig.\ref{FigOrbitNet}), $P_x$ is the conserved generalized momentum projection. Here $b^2_H = e \hbar H$ is the elementary "magnetic area" in the momentum space, related to the "magnetic (cyclotron) energy" as $\hbar \omega_H=b^2_H / m$. 

The semiclassical solution of  Eq.(\ref{LOSchrod}) is  
%
\begin{eqnarray}
G_\alpha(p_y)=\frac{A_\alpha}{\sqrt{v_x^{\alpha}}}e^{i(P_x p_y/b_H^2)}e^{iS_\alpha(p_y)/b_H^2},
\label{LOWF}
\end{eqnarray}
%
where $A_\alpha$ are constants which are determined by matching of the wave functions. The velocity
%
\begin{eqnarray}
v_x^{\alpha}(p_y) = \frac{\partial \varepsilon_\alpha (\mathbf{p})}{\partial p_x}\Big|_{p_x= d S_\alpha/dp_y}
\label{velocity_x}
\end{eqnarray}
%
is determined by the momentum space area (phase)
%
\begin{eqnarray}
S_{\alpha}(p_y) =\int_0^{p_y} p_x^{\alpha}(\varepsilon, p_y^{\prime})d p_y^{\prime},
\label{phaseS}
\end{eqnarray}
%
where the integrated function $p_x^{\alpha}(\varepsilon, p_y^{\prime})$ is found from the algebraic equation
%
\begin{eqnarray}
\varepsilon_\alpha(p_x,p_y^\prime)=\varepsilon.
\label{pxequation}
\end{eqnarray}
%
The beginning of the integration $p_y=0$ is chosen to be a starting point of the quasi-particle motion along the trajectory $\alpha$ (the arrows in Fig.\ref{FigOrbitNet} show the direction of quasi-particle motion). 

Inside the small areas (shaded and marked by thick dots in the figure), in which the 
semiclassical trajectories closely approach each other, the quasi-particle undergoes a quantum tunnelling between them (so-called magnetic breakdown \cite{Falikov,Blount} - MB). At the MB points, the outgoing and incoming wave functions are related by the $2 \times 2$ unitary matrix - the "MB matrix"
%
\begin{eqnarray}
\hat{\tau}
=e^{i \chi}
\left(
\begin{array}{cc}
t & r\\
-r^{\star} & t^{\star}
\end{array}
\right),
\label{MBmatrix}
\end{eqnarray}
%
where $t$ and $r$ are the complex amplitudes of the before-mentioned probabilities for quasi-particle motion along the diamond-shaped or circular orbit, respectively, after scattering in the MB region, fulfilling the condition $|t|^2+|r|^2=1$, $\chi$ is the real phase. All four independent parameters of the MB matrix depend on magnetic field \cite{MBpobabTouch1,MBpobabTouch2,Vorontsov,Fortin}.
Therefore, in the momentum representation, the wave  function of the quasi-particle  moving along the 2D net under magnetic field takes the form
%
\begin{eqnarray}
G(p_x,p_y)=e^{i(P_x p_y/b_H^2)}\sum_{\alpha}\frac{A_\alpha}{\sqrt{v_x^\alpha}}e^{iS_\alpha(p_y)/b_H^2}
\label{totalWF}
\end{eqnarray}
%
where the wave functions under summation are coupled by the MB matrix Eq.(\ref{MBmatrix}). Note that the dependence of the wave function $G(p_x,p_y)$ on $p_x$  is implicitly present in the $A_\alpha$ and the position of the trajectory $\alpha$.

Matching all the semiclassical wave functions, Eq.(\ref{LOWF}), related to the first Brilliouin  zone with the MB matrix, one finds a set of 8 algebraic equations for coefficients $A_\alpha$:
%
\begin{eqnarray}
\left(
\begin{array}{c}
A_1 \\
A_5\\
\end{array}
\right)
 =e^{i \chi}
 \left(
\begin{array}{cc}
t & r\\
-r^{\star} & t^{\star}
\end{array}
\right)
\left(
\begin{array}{c}
A_2 e^{i\Theta_2} \\
A_4 e^{i\Theta_4}\\
\end{array}
\right) \nonumber
\end{eqnarray}
\begin{eqnarray}
\left(
\begin{array}{c}
A_4 \\
A_{10}\\
\end{array}
\right)
 =e^{i \chi}
 \left(
\begin{array}{cc}
t & r\\
-r^{\star} & t^{\star}
\end{array}
\right)
\left(
\begin{array}{c}
A_9 e^{i\Theta_9} \\
A_3 e^{i\Theta_3}\\
\end{array}
\right)\nonumber
\end{eqnarray}
\begin{eqnarray}
\left(
\begin{array}{c}
A_9 \\
A_{12}\\
\end{array}
\right)
 =e^{i \chi}
 \left(
\begin{array}{cc}
t & r\\
-r^{\star} & t^{\star}
\end{array}
\right)
\left(
\begin{array}{c}
A_8 e^{i\Theta_8} \\
A_{11} e^{i\Theta_{11}}\\
\end{array}
\right)\nonumber
\end{eqnarray}
\begin{eqnarray}
\left(
\begin{array}{c}
A_6 \\
A_8\\
\end{array}
\right)
 =e^{i \chi}
 \left(
\begin{array}{cc}
t & r\\
-r^{\star} & t^{\star}
\end{array}
\right)
\left(
\begin{array}{c}
A_7e^{i\Theta_7} \\
A_5 e^{i\Theta_5}\\
\end{array}
\right),
\label{Match}
\end{eqnarray}
%
where 
%
\begin{eqnarray}
\Theta_{\alpha} =\frac{S_\alpha(p_{\alpha}^e)}{b_H^2}
 \label{PhaseGain}
\end{eqnarray}
%
are the phase gains obtained by the wave function Eq.(\ref{LOWF}) during the quasi-particle motion from the beginning to the end of the trajectory $\alpha$. There, $p_{\alpha}^{e}$ is the $p_y$-coordinate of the trajectory ending point with respect to the direction of motion.

As one sees from the Eq.(\ref{totalWF}), the sought wave function $G(p_x,p_y)$ is  the proper function of the translation operator
$\hat{T}_y= e^{b_y\partial/\partial p_y}$, i.e.
%
\begin{eqnarray}
\hat{T}_yG(p_x,p_y)=e^{i(P_xb_y/b_H^2)}G(p_x,p_y)
\label{TranslOperp_y}
\end{eqnarray}
%
for any value of magnetic field, where $b_y$ is the (reciprocal) lattice constant in $p_y$-direction of the reciprocal (momentum) space after reconstruction. The eigenvalue is characterized by the proper generalized momentum $P_x$.
The 2D net of the MB-coupled trajectories is geometrically periodic (see Fig.\ref{FigOrbitNet}) in both $p_x$- and $p_y$-directions and, hence, one may assume that the quasi-particle wave function $G(p_x, p_y)$ is a proper function not only of the translation operator $\hat{T}_y$, but also of $\hat{T}_x= e^{b_x\partial/\partial p_x}$ with the proper generalized momentum $P_y$, i.e.
%
\begin{eqnarray}
\hat{T}_x G(p_x,p_y) = e^{i(P_y b_x/b_H^2)} G(p_x,p_y),
\label{TranslOperp_x}
\end{eqnarray}
%
where $b_x$ is the analogous reciprocal lattice constant in $p_x$-direction.
In this case, as it follows from  Eqs.(\ref{LOWF},\ref{TranslOperp_y},\ref{TranslOperp_x}) and Fig.\ref{FigOrbitNet}, 
constants $A_\alpha$ are additionally coupled by the conditions
%
\begin{eqnarray}
A_2= A_8 e^{iP_y b_x/b_H^2}; \;\; A_{11} = A_4 e^{-iP_y b_x/b_H^2}; \nonumber \\ 
A_1= A_9 e^{iP_y b_x/b_H^2}; \;\; A_{12} = A_5 e^{-iP_y b_x/b_H^2}; \nonumber \\ 
A_6= A_4 e^{iP_x b_y/b_H^2}; \;\;\; A_{3} = A_5 e^{-iP_x b_y/b_H^2}; \nonumber \\ 
A_7= A_9 e^{iP_x b_y/b_H^2}; \;\; A_{10} = A_8 e^{-iP_x b_y/b_H^2},
\label{BlochCondition}
\end{eqnarray}
%
where we note the phase factors determined by the conserved generalized momenta $P_x$ and $P_y$.
As  a result, we have 16 equations, Eq.(\ref{Match}) and (\ref{BlochCondition}) for 12 unknown coefficients $A_{\alpha}$. Since  the number of equation is larger than the number of unknowns, the system is overdetermined and, hence, cannot be solved as such.\   

However, as one can see from the definition of the phase gains Eq.(\ref{PhaseGain}) and Fig.\ref{FigOrbitNet}, there are following relations between phases:  
%
\begin{eqnarray}
\Theta_2=\Theta_8 +\frac{b_x b_y}{2 b^2_H}; \hspace{0.3cm} \Theta_3=\Theta_5;  \nonumber \\
\Theta_{11}=\Theta_4 -\frac{b_x b_y}{2 b^2_H}; \hspace{0.3cm} \Theta_{7}=\Theta_9.
\label{PhaseRelations}
\end{eqnarray}
%
If magnetic field is chosen to satisfy condition
%
\begin{eqnarray}
\frac{b_x b_y}{2 b^2_H} = \pi \frac{\Phi_0}{a_x a_y H}= 2\pi l,
\label{ZakCond1}
\end{eqnarray}
%
where $l$ is an integer (in our case $l \gg 1$), $a_x$ and $a_y$ are reconstructed lattice constants in the real space and $\Phi=a_x a_y H$ is thus magnetic flux piercing the unit cell, only 4 equations from the set of 8 equations Eqs.(\ref{Match}) are independent, hence making the total system of equations solvable \cite{comment}. As already mentioned in Introduction, this condition is consistent with works of Zak and Hofstadter \cite{Zak,Hofstadter}, so we call it the "Zak's condition" in this paper. Inserting Eq.(\ref{PhaseRelations}) into  Eq.(\ref{Match}) one easily finds that the first 4 and the last 4 equations are identical.
Therefore, for the magnetic fields satisfying the "Zak's condition" Eq.(\ref{ZakCond1}), the set of equations Eq.(\ref{Match}) reduces to the system of four equations for four unknowns, i.e.
%
\begin{eqnarray}
\begin{array}{c}
+r A_4 e^{i (\Theta_{11}-K_y)}+ t A_8 e^{i \Theta_{8}}-
A_9 e^{-i \chi }=0; \;  \\
-t^\star A_5 e^{i \Theta_{5}}+  A_8 e^{-i \chi} +
r^\star A_9e^{i( \Theta_{7}+K_x)}=0;\\
-t^\star A_4 e^{i \Theta_{4}}+ A_5 e^{-i \chi}+
r^\star A_8 e^{i (\Theta_2 +K_y) }=0;\\
- A_4 e^{-i\chi}+ r A_5 e^{i (\Theta_3 -K_x)} +
t A_9 e^{i( \Theta_{9})}=0,
\end{array}
\label{homogenset}
\end{eqnarray}
%
that determine the quasi-particle wave function Eq.(\ref{totalWF}) and the quasi-particle spectrum.\

Calculating the determinant of the above set of homogeneous algebraic equations
%
\begin{eqnarray}
D(\varepsilon, \mathbf{P})&=&\sin\left[\frac{ S_{\diamond}(\varepsilon)}{2 b_H^2}+ 2 \chi \right] + \nonumber \\ 
|t| |r| &\times & \left\{ \sin \left[\frac{b_yP_x}{ b_H^2} + \mu - \eta \right] + \sin\left[\frac{b_xP_y}{ b_H^2} - \mu - \eta \right]
\right \}, \nonumber \\
 \label{determinant}
\end{eqnarray}
%
where $\mathbf{P}=(P_x,P_y)$, one finds the dispersion equation for quasi-particles moving in the 2D lattice under magnetic field
%
\begin{eqnarray}
D(\varepsilon, \mathbf{P})=0.
 \label{dispersionequation}
\end{eqnarray}
%
Here $S_{\diamond}$ is the area of the diamond-shaped trajectory in Fig.\ref{FigOrbitNet}, related to the circular-shaped one $S_{\circ}=2\pi m \varepsilon$, by the relation $S_{\diamond}(\varepsilon)= b_x b_y - S_{\circ} = b_x b_y -2\pi m \varepsilon$, while $\mu$ and $\eta$ are phases of the probability amplitudes $t$ and $r$, respectively.\

\begin{widetext}

\begin{figure}
\centerline{\includegraphics[width=12.0cm]{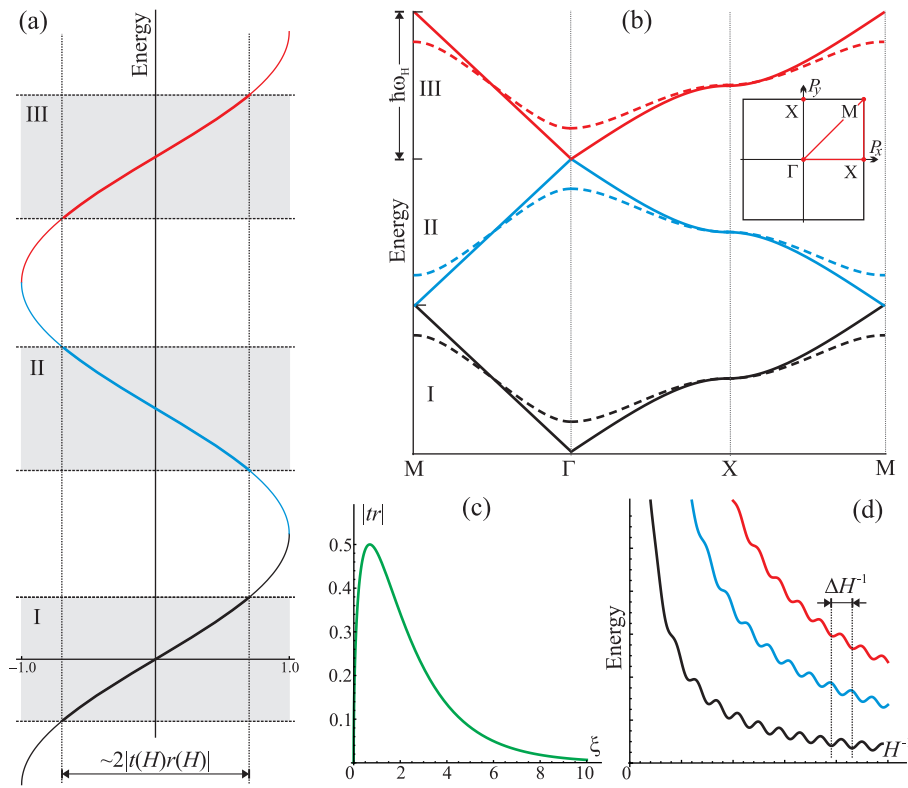}}
\caption {Spectrum of quasi-particles: (a) Formation of the energy bands (along vertical axis, indexed by I, II, III,...) depending on argument of the $\arcsin$ function (curve) in Eq.(\ref{spectrum1}) shown on the horizontal axis, which is proportional to $|t(H)r(H)|$, determining the width of the bands Eq.(\ref{bandwidth}). (b) Energy dispersion in the momentum space along the lines between the characteristic points (see the inset). Width of the magnetic energy bands is maximal, i.e. $\hbar \omega_H$, for $|tr|=1/2$ (full lines) and the gap is closed. For  $|tr| < 1/2$ (dashed), the band-width is smaller and gaps between bands appear. (c) Function $|t(\xi)r(\xi)|=\exp{[-\xi/2]}\sqrt{1-\exp{[-\xi]}}$, where the argument $\xi \equiv \Delta^2 (\hbar \omega_H)^{-4/3}\varepsilon_F^{-2/3}$ depends on magnetic field. (d) Energy spectrum taken at the $\Gamma$-point depending on magnetic field exhibits rapid oscillations with period $\Delta H^{-1} \approx h e / p_F^2$ with respect to the inverse magnetic field $H^{-1}$ ($p_F$ is the Fermi momentum determined by the area of the pocket encircled by the quasi-particle trajectory, e.g. $S_{\circ} \approx \pi p_F^2$). For simplicity, we set the phase $\chi=\pi/4$ in Eq.(\ref{spectrum1}), valid in the limit of Landau levels.}
\label{figspectrum1}
\end{figure}

\end{widetext}

The quasi-particle spectrum found from Eq.(\ref{dispersionequation}) reads
%
\begin{eqnarray}
\varepsilon_s(P_x,P_y) &=& \hbar \omega_H \Big\{ s+\frac{2\chi}{\pi} \nonumber \\
&+& \frac{(-1)^s}{\pi} 
 \arcsin \Big[ |tr| 
\left( \sin\left(\frac{b_y P_x}{ b_H^2} + \mu - \eta \right) \right. \nonumber \\
 &+& \left. \sin\left(\frac{b_x P_y}{ b_H^2} - \mu - \eta \right) \right)
\Big ]\Big\},\;\;\;\;
\label{spectrum1}
\end{eqnarray}
%
where $s=0, 1, 2, ...$ is the new band index, and is shown in Fig.\ref{figspectrum1}.
%
One can see from Eq.(\ref{spectrum1}) and Fig.\ref{figspectrum1}(a,b) that the spectrum consists of the periodic series of alternating (magnetic) energy bands of the width
%
\begin{eqnarray}
W(H)=\frac{2}{\pi}\arcsin[2|t(H)r(H)|] \, \hbar \omega_H
\label{bandwidth}
\end{eqnarray}
%
and gaps of the width $\hbar \omega_H - W$ between them.
The function $|t(H)r(H)|$ takes values between 0 and 1/2 (see Fig.\ref{figspectrum1}(c)). In the limit $|t(H)|=0$, or $|r(H)|=0$ we have Landau levels due to the Landau quantization of circular, or diamond-shaped orbits, respectively. In the intermediate regime, starting with $|t|\ll 1$ in one and $|r|\ll 1$ in another mentioned limit, the Landau levels are broadened into magnetic energy bands due to the MB, with gaps in between. In the limit $|tr|=1/2$, the gap closes in the $\Gamma$- and $M$-point in the Brillouin zone, with approximately linear dispersion around them (see Fig.\ref{figspectrum1}(b)). Spectrum also exhibits rapid oscillations with respect to magnetic field (see Fig.\ref{figspectrum1}(d)) with period determined by the area of the pocket encircled by the semiclassical trajectory.\ 

It is worth noticing that the quasi-particle group velocity is of the order of the Fermi velocity $v_F$ in spite of the fact that the band-width is very narrow, i.e. $\mathbf{v} =\partial \varepsilon_s/\partial \mathbf{P} \sim |t r| b/m \sim 10^6 m/s$ 
for $|tr| \sim 1$, where, for the sake of simplicity, we assumed the cubic lattice symmetry, i.e. $b_x =b_y \equiv b$.\

Also, it is important to emphasize, once more, that the spectrum Eq.(\ref{spectrum1}), as well as all quantities following from it (e.g. conductivity discussed in the next section) is valid for the magnetic fields satisfying the "Zak's condition" Eq.(\ref{ZakCond1}). Number $l$ determines how many unit cells in real space "host" one flux quantum, and, for materials with lattice constant of Angstr\"om size, this number is huge, of the order of $10^4$ for magnetic fields of the order of 10T. Resolution of magnetic field, corresponding to the change of $l$ by one, is of the order of $10^{-3}$T in such systems.  It is known \cite{Solyom} that the "Zak / Hofstdater effect" of flux quantization vs. the underlying lattice is observable for specially engineered structures with large unit cells, very strong magnetic fields and clean samples. It is also possible to generalize the present analysis to the periodicities in the reciprocal space involving multiple reciprocal unit cells, e.g. by using the transfer matrix or some other method, however, neither of the two above-mentioned issues are not in the focus of this work and will be discussed elsewhere.\\

\section{Kinetics of quasi-particles in the 2D net of semiclassical trajectories  under the MB conditions \label{SecKinetics}}

In this section we consider kinetics of quasi-particles in the 2D MB net under magnetic field which satisfies the "Zak's condition" Eq.(\ref{ZakCond1}). Due to the MB scattering, for which the corresponding amplitudes $|t(H)|$ and $|r(H)|$ are comparable, a 2D quasi-particle is delocalized in both directions, moving along the diamond-shaped and circular orbits as well as tunnelling between them, having two conserved generalized momenta, $P_x$ and $P_y$, consequently resulting in the quasi-particle dispersion equation Eq.(\ref{dispersionequation}).
The width of the quasi-particle energy band, Eq.(\ref{bandwidth}), essentially depends on the function $|t(H)r(H)|$, thus defining two characteristic limits of weak and strong magnetic field in which $|t(H)r(H)| \rightarrow 0$ (see \ref{figspectrum1}(c)). Consequently, $W \rightarrow 0$ in those limits, determining two intervals of  magnetic field in which kinetics (and dynamics) of quasi-particles is qualitatively different:\

1) If the band-width is much greater than the energy uncertainty due to scattering on impurities, i.e. $W \gg \hbar \nu_0$, the quasi-particle is delocalized  and its group velocity is large in both directions, i.e. $\mathbf{v}=\partial \varepsilon_s/\partial 
 \mathbf{P}\sim |t(H)r(H)| v_F$.\
 
2) If the band-width is much smaller than the energy uncertainty due to scattering on impurities, i.e. $W \ll \hbar \nu_0$, the effective spectrum is a  serial of Landau energy levels corresponding to the quasi-particle motion along the diamond-shaped trajectory for $t(H) \rightarrow 1$, or the circular one for $r(H) \rightarrow 1$ (see Fig.\ref{FigOrbitNet}). Therefore, in those limiting cases the magnetoconductivity tensor is the standard one, corresponding to the closed Fermi surfaces, well known from the literature \cite{LAK}.\\

\textbf{1.} \textbf{Magnetoconductivity of quasi-particles delocalized by the MB in both $x$- and $y$-direction of the 2D net}
\bigskip

Here we assume the inequality $W \gg  \hbar \nu_0$ and find the magnetoconductivity tensor for quasi-particles. 

The linearized equation for the density matrix $\hat{\rho} =f_0(\hat{H}_0)+\hat{\rho}^{(1)}$ for quasi-particles under electric field $\mathbf{E}$ reads
%
%
\begin{eqnarray}
\frac{1}{i \hbar}\left[\hat{H}_0,\hat{\rho}^{(1)} \right]+ \frac{\hat{\rho}^{(1)}}{t_0}=-\frac{e \mathbf{E}}{i \hbar}
\left[ \hat{\mathbf{r}},f_0(\hat{H}_0) \right].
 \label{DensityMatrixEquat}
\end{eqnarray}
%
Here $\hat{H}_0$ is the quasi-particle effective Hamiltonian, the Schr\"odinger equation for which is
%
\begin{eqnarray}
\hat{H}_0 |n,\mathbf{P}\rangle =\varepsilon_n(\mathbf{P})|n,\mathbf{P}\rangle,
\label{SchroedingEquat}
\end{eqnarray}
%
where the proper energy $\varepsilon_n(\mathbf{P})$ is defined by the dispersion equation Eq.(\ref{dispersionequation}) and the proper functions are Bloch functions, $t_0$ is the relaxation time (time of the mean free path of a quasi-particle), $f_0$ is the Fermi distribution function, and $\hat{\rho}^{(1)}$ is the first correction to $\hat{\rho}$ with respect to linearization.\

Taking matrix elements of the density matrix equation Eq.(\ref{DensityMatrixEquat}) and using the fact that both
projections of velocity are finite, $v_x \neq 0$ and $v_y \neq 0$, one finds the density matrix
%
\begin{eqnarray}
\rho_{\kappa \kappa^\prime}^{(1)} =e \mathcal{\bf{E}}
\frac{i \hbar  \mathbf{v}_{\kappa \kappa^\prime}}{\varepsilon_\kappa-\varepsilon_{\kappa^{\prime}} +i\hbar \nu_0}\frac{f_0(\varepsilon_\kappa)-f_0(\varepsilon_{\kappa^{\prime}})}{\varepsilon_\kappa-\varepsilon_{\kappa^{\prime}}},
 \label{DensityMatrix}
\end{eqnarray}
%
where $\kappa =(\mathbf{P}, n)$ and $\mathbf{v}_{\kappa \kappa^\prime}$ are the the velocity matrices
%
\begin{eqnarray}
\mathbf{v}_{\kappa \kappa^\prime}=\frac{\partial \varepsilon_\kappa }{\partial \mathbf{P}}\delta(\mathbf{P}-\mathbf{P}^\prime)\delta_{n,n^\prime}.
 \label{velocityME}
\end{eqnarray}
%

As the quasi-particle group velocity is not zero, the main contribution to the current $\mathbf{j} =-2e \mathrm{Tr}(\hat{\mathbf{v}}\hat{\rho})$, where factor 2 accounts for the quasi-particle spin and $\mathrm{Tr}$ denotes the trace operation, comes from the diagonal element of the density matrix Eq.(\ref{DensityMatrix}) and, hence, the conductivity is 
%
\begin{eqnarray}
\sigma_{ik} =-2 e^2 t_0 \sum_n \int \frac{d \mathbf{P}}{(2 \pi \hbar)^2} \frac{\partial
 \varepsilon_n}{\partial P_i}  
 \frac{\partial \varepsilon_n}{\partial P_k} 
\frac{\partial f_0}{\partial \varepsilon}\Big|_{\varepsilon=\varepsilon_n(\mathbf{P})},
 \label{sigma}
\end{eqnarray}
%
where $i,k \in \{x,y\}$ (details of calculations are presented in the Appendix \ref{AppCurrent}).\

For calculations of various thermodynamic and kinetic coefficients of quasi-particles with complicated dispersion laws, it is convenient to use the approach developed by Slutskin \cite{Slutskin1} for electrons under the MB in which one uses the dispersion equation for calculations instead of the complicated dispersion law.  
Differentiating dispersion equation Eq.(\ref{dispersionequation}) with respect to $\mathbf{P}$, i.e. $\partial_{\varepsilon} D \cdot \partial_{P_{i}} \varepsilon + \partial_{P_{i}}D = 0$, one finds
%
\begin{eqnarray}
v_i=\frac{\partial
\varepsilon_n}{\partial P_i}  = - \frac{\partial D(\varepsilon, \mathbf{P})/\partial P_i}{\partial D(\varepsilon, \mathbf{P})/\partial \varepsilon},  
\label{velocityD}
\end{eqnarray}
%
with $D\left(\varepsilon,\mathbf{ P} \right)=0$.
This expression allows to present the conductance, Eq.(\ref{sigma}), in the sought form (see Appendix \ref{AppDcurrent})
%
\begin{eqnarray}
\sigma_{ik} =  
- 2 e^2 t_0 \int d\varepsilon \frac{\partial f_0}{\partial \varepsilon} \int{ \frac{d \mathbf{P}}{(2 \pi \hbar)^2} \frac{\frac{\partial
D}{\partial P_i} \frac{\partial
D}{\partial P_k}}{|\frac{\partial
D}{\partial \varepsilon}|} \delta\left[D(\varepsilon,\mathbf{ P})\right]}. \nonumber\\ 
 \label{sigma1}
\end{eqnarray}
%
Using  Eq.(\ref{dispersionequation}), one can write Eq.(\ref{sigma1}) in the form
%
\begin{eqnarray}
\sigma_{ik} &=& -2 e^2 t_0\frac{\bar{b}_i \bar{b}_k}{\pi m b_H^2}|t r|^2 
\int d\varepsilon
\frac{\partial f_0/\partial \varepsilon}{|\cos \Theta_{\diamond}|}
\int  \frac{d \mathbf{P}}{(2 \pi \hbar)^2}  \nonumber \\
&\times & \cos\Theta_i \cos\Theta_k  
\delta \left[\sin \Theta_{\diamond}(\varepsilon) -|r t|\left(   \sin\Theta_x  + \sin\Theta_y   \right) \right], \nonumber \\
 \label{sigma3}
\end{eqnarray}
%
where bar denotes the index conjugation, i.e. $\bar{b}_x =b_y$,  $\bar{b}_y =b_x$, and
%
\begin{eqnarray}
\Theta_{\diamond} &=& \frac{S_{\diamond}(\varepsilon)}{2 b^2_H} +\chi; \nonumber \\
\Theta_x &=& \frac{P_x b_y}{b_H^2}  +\mu-\nu; \nonumber \\
\Theta_y &=& \frac{P_y b_x}{b_H^2} -\mu-\nu.
\label{sigmaPhases}
\end{eqnarray}
%
One sees that the integrand in Eq.(\ref{sigma3}),
%
\begin{eqnarray}
Q = \cos\Theta_i \cos\Theta_k  \delta \left[\sin \Theta_{\diamond}(\varepsilon) -|tr|\left(   \sin\Theta_x + \sin\Theta_y   \right) \right],\nonumber\\
 \label{Integrand}
\end{eqnarray}
%
is a  $2 \pi$-periodic function of $\Theta_{x}$ and $\Theta_{y}$, provided they are considered as free variables, which allows to expand the the integrand in the double-Fourier series
%
\begin{eqnarray}
Q_{ik}=\sum_{l_1=-\infty}^{\infty}\sum_{l_2=-\infty}^{\infty}A_{l_1 l_2}^{ik}e^{i(l_1 \Theta_x +l_2 \Theta_y)}.
 \label{Fourier}
\end{eqnarray}
%
Since $\varepsilon_F/\hbar \omega_H \gg 1 $,  the exponents are fast-oscillating functions 
of $P_x$ and $P_y$ and, hence, the main contribution to the integral in Eq.(\ref{sigma3}) comes from the Fourier factor with $l_1 =l_2=0$, and therefore the sought integrand is approximately
%
\begin{eqnarray}
Q_{ik} \approx A_{00}^{ik}  = 
\frac{1}{(2 \pi)^2}\int_{-\pi}^{\pi}d \varphi_x \int_0^{2 \pi}d \varphi_y 
\cos \varphi_i\cos \varphi_k  \nonumber \\
\times \delta \left[\sin \Theta_{\diamond}(\varepsilon) -|tr|\left(   \sin\varphi_x + \sin\varphi_y   \right) \right].
 \label{Q0}
\end{eqnarray}
%

\begin{widetext}

\begin{figure}
\centerline{\includegraphics [width=12.0cm] {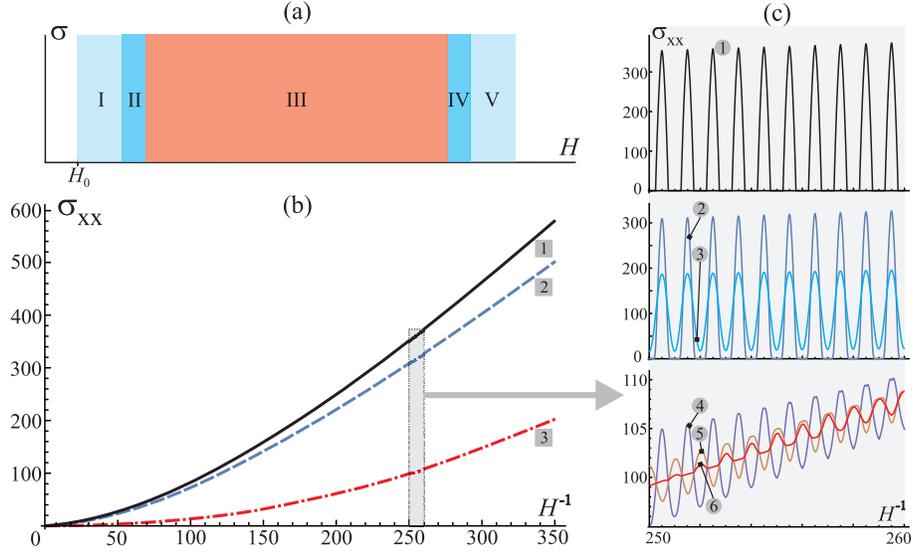}}
\caption {(a) Characteristic intervals of magnetic field with qualitatively different behaviour of magnetoconductivity with respect to development of the MB. $H_0$ is the lower limit of magnetic field in our consideration, for which $t_0\omega_{H_{0}} \gg 1$ holds ($t_0$ is the quasi-particle mean free path time). In regions I and V we have either small, or large magnetic field for which $|r| \rightarrow 0$, or $|t| \rightarrow 0$, so the quasi-particles move along diamond-shaped or circular closed orbits, respectively. The magnetoconductivity attains the standard textbook form for closed hole-like, or electron-like orbits, i.e. Eq.(\ref{sigmaconventionalh}) or Eq.(\ref{sigmaconventionale}), respectively. In region III, $|t|$ and $|r|$ are comparable, magnetoconductivity is given by Eq.(\ref{sigmafinal}), and depicted at panels (b) and (c), here scaled by the factor multiplying the integral in Eq.(\ref{sigmafinal}), proportional to $H^{-1}$ through $b_H^2$, which has dimension of the 2D conductivity. II and IV are the crossover regions between the regimes described above. (b) An envelope (upper) of the oscillatory part of the longitudinal magnetoconductivity in the region III versus inverse magnetic field (expressed in $\varepsilon_F/\hbar \omega_H$ dimensionelss units), depending on temperature: $k_BT/\varepsilon_F=$0; 0.0001; 0.01 for curve 1; 2; and 3 respectively. (c) An inset from panel (b), showing the oscillations of longitudinal magnetoconductivity versus inverse magnetic field (expressed in $\varepsilon_F/\hbar \omega_H$ dimensionelss units), depending on temperature: $k_BT/\varepsilon_F=$0; 0.0001; 0.0005; 0.001; 0.005; 0.01 for curve 1; 2; 3; 4; and 5, respectively.}
\label{FigSigmaxx}
\end{figure}

\end{widetext}

Inserting $Q_{ik}$ in the integral in  Eq.(\ref{sigma3}) one finds the conductivity in the form
%
\begin{widetext}
\begin{eqnarray}
\sigma_{ik} =-2 e^2 t_0\frac{\bar{b}_i \bar{b}_k |t r|^2}{\pi m b_H^2} \frac{b_x b_y}{(2\pi 
\hbar)^2}
\int d\varepsilon
\frac{\partial f_0/\partial \varepsilon}{|\cos \Theta_{\diamond}|}
\int_{-\pi}^{ \pi}\int_{-\pi}^{ \pi} 
d \varphi_x   
d \varphi_y
\sin \varphi_i\sin \varphi_k  
\, \delta \left[\sin \Theta_{\diamond}(\varepsilon) -|tr|\left(   \cos\varphi_x + \cos\varphi_y   \right) \right], 
 \label{sigma4}
\end{eqnarray}
\end{widetext}
%
where, while writing this expression, we shifted the integration variables $\varphi_{x,y} \rightarrow
\varphi_{x,y} +\pi/2$ and, hence, the limits of integration in order to explicitly  match the integral to the symmetry of the system.  
Using the symmetric properties of the integral and integrating the delta-function we obtain components of the magnetoconductivity tensor:
%
\begin{widetext}
\begin{eqnarray}
\sigma_{xx} &=& -e^2 t_0 \frac{b_y^2}{\pi^4 m \hbar^2}\left[\frac{1}{2\pi} \left(\frac{b_x b_y}{2 b_H^2}\right)\right]
\int d\varepsilon
\frac{\partial f_0/\partial \varepsilon}{|\cos \Theta_{\diamond}|}
 \int_{-\pi}^{+\pi}d \varphi \sqrt{|tr|^2-\Big(|tr| \cos\varphi-\sin\Theta_{\diamond}(\varepsilon)\Big)^2} \nonumber \\
&\times & \mathbf{\Theta}\left[  |tr|^2-\Big(|tr| \cos\varphi-\sin\Theta_{\diamond}(\varepsilon)\Big)^2 \right]; \nonumber \\
\sigma_{xy}&=&0,
 \label{sigmafinal}
\end{eqnarray}
\end{widetext}
%
where $\mathbf{\Theta}$ is the Heaviside unit-step function, while the other two components, $\sigma_{yy}$ and $\sigma_{yx}$, are simply obtained from the above expression by changing the indices $x \leftrightarrow y$. One should note the factor appearing in the square brackets multiplying the integral, which equals the integer number determined by the condition Eq.(\ref{ZakCond1}): it is proportional to $1/H$ (through $b_H^2$) that appears in the discrete quantities.
Therefore, the obtained equation for the conductivity tensor, Eq.(\ref{sigmafinal}), is valid at discrete values of the magnetic field $H_l$ determined by the Eq.(\ref{ZakCond1}). From the latter equation and Eq.(\ref{sigmafinal}) one easily finds that the distance between the neighbouring points $H^{-1}_{l} - H^{-1}_{l+1}$ and the period of conductance oscillations, $\Delta H^{-1}$, are incommensurate. From here it follows that the points $H^{-1}_{l}$ gathered inside one period of the conductance oscillations cover it everywhere  densely and, hence, the equation Eq.(\ref{sigmafinal}) presents the conductivity tensor with very fine resolution. 
This result is valid in the limit of (magnetic) band-width significantly larger than the energy uncertainty related to quasi-particle scattering on impurities, i.e. $W \gg \hbar \nu_0$. It is depicted in Fig.\ref{FigSigmaxx}, and is relevant for interval of magnetic fields marked by III in panel (a), where the characteristic intervals  of magnetic field with qualitatively different behaviour of magnetoconductivity with respect to the MB development are explained; for the sake of simplicity the phase in the integral in Eqs.(\ref{sigmaPhases},\ref{sigmafinal}) is taken  $\chi=\pi/4$ .\

There are several important points to notice regarding the result Eq.(\ref{sigmafinal}):\

First of all, a remarkable fact is that the Hall effect is absent, i.e. $\sigma_{xy}=\sigma_{yx} =0$. It is easy to obtain that result simply from the symmetry of the integral Eq.(\ref{sigma4}) by changing $\varphi_i \rightarrow - \varphi_i$. It happens in a wide range of magnetic fields in which the MB gives comparable probabilities $|t|^2$ and $|r|^2$ for motion along the closed orbits and tunneling between them. This means that the quasi-particle is delocalized in both directions of the 2D net, by the MB, performing motion as if there is effectively no magnetic field acting on it. Below we show that the Hall effect restores in the limits  of relatively weak and relatively strong magnetic fields when quasi-particles move along the closed semiclassical trajectories.\

The diagonal magnetoconductivity exhibits oscillations under a change of magnetic field - Fig.\ref{FigSigmaxx}(b) shows the envelope, while the inset drawn on panel (c) shows the oscillations.
As one can see from Eq.(\ref{sigmafinal}) and Fig.\ref{FigSigmaxx}, the period of oscillations 
is close to the period of the conventional de Haas-van Alphen or Subnikov-de Haas oscillations, but their amplitude is giant, becoming less pronounced with increasing temperature as well as changing the phase (due to integration of components with more periods depending on energy in the "$k_B T$-window" around the Fermi energy).\

%
%

\textbf{2. Magnetoconductivity in the limit of narrow band-width}
\bigskip

As noted above, if the MB probability amplitudes are such that the inequality $W \ll \hbar \nu_0 \ll \hbar \omega_H$ is fulfilled, i.e. when the (magnetic) band-width is the smallest parameter, the quasi-particle band structure  is a set of Landau levels slightly broadened by the MB tunneling between the closed orbits. In this limit we neglect that level broadening, i.e. we consider quasi-particles moving semiclassically along either the diamond-shaped orbit (rather weak magnetic fields, $|r| \ll 1$), or along the circular orbits (strong magnetic fields, $|t| \ll 1$) shown in Fig.\ref{FigOrbitNet}. As a result, the magnetoconductivity tensor is the conventional one \cite{LAK}, that is:\\

a) $|t| \rightarrow 1$ for the hole-like trajectories
\begin{eqnarray}
\sigma^{(h)}_{xx}=\sigma^{(h)}_{yy} &=& \frac{\sigma_0^{(h)}}{(t_0 \omega_H)^2} ; \nonumber\\
\sigma^{(h)}_{xy}= -\sigma^{(h)}_{yx} &=& \frac{n^{(h)}e}{H},
 \label{sigmaconventionalh}
\end{eqnarray}
%

b) $|t| \rightarrow 0$ for the electron-like trajectories
%
\begin{eqnarray}
\sigma^{(e)}_{xx}=\sigma^{(e)}_{yy} &=&\frac{\sigma_0^{(e)}}{(t_0 \omega_H)^2}; \nonumber\\
\sigma^{(e)}_{xy}= -\sigma^{(e)}_{yx} &=& -\frac{n^{(e)}e}{H},
\label{sigmaconventionale}
\end{eqnarray}
%
where $\sigma_{0}^{(e,h)}= n^{(e,h)}e^2 t_0/m$  is the conductivity of electrons $(e)$, or holes $(h)$ in the absence of magnetic field, while $n^{(e)} =\pi p^2_F/(2\pi \hbar)^2$  is the  concentration of electrons, and  $n^{(h)} =[b^2 -\pi p^2_F]/(2\pi \hbar)^2$  concentration of holes. Here, for the sake of simplicity, we consider $b_x=b_y \equiv b$ and equal effective masses of electrons and holes, i.e. $m_e=m_h \equiv m$.\\

\textbf{3. The Hall coefficient}
\bigskip

\begin{figure}
\centerline{\includegraphics[width=6.5cm]{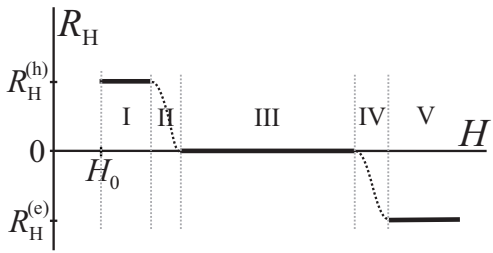}}
\caption {Dependence of the Hall coefficient $R_H$ on magnetic field (through the MB probability amplitude $|t(H)|$). In respectively weak magnetic field (still greater than $H_0$ for which $t_0\omega_{H_{0}} \gg 1$) - interval I, as well as in strong enough fields - interval V, such that the quasi-particle band-widths are negligible, i.e. $W \ll  \hbar\nu_0 \ll \hbar\omega_H$, the quasi-particles move along the diamond-shaped (hole) and circular (electron) trajectories, respectively. Consequently, the Hall coefficient is given by Eq.(\ref{Hallcoefficienth}), or Eq.(\ref{Hallcoefficiente}), respectively. In the wide intermediate magnetic field range in which $W \gg \hbar\nu_0 $ (interval III), the quasi-particles freely move in both $x$- and $y$-direction resulting in $R_H =0$ due to the reciprocal lattice symmetry. Intervals II and IV are the crossover regions between the above-mentioned ones.}
\label{FigHall}
\end{figure}

Behaviour of the Hall coefficient
%
\begin{eqnarray}
R_H \equiv \frac{E_y}{j_x H}=\frac{\sigma_{xy}}{(\sigma_{xx}\sigma_{yy}+\sigma_{xy}^2)H}
\label{HallcoefficientGeneral}
\end{eqnarray}
%
with a change of magnetic field in the system under consideration is particularly remarkable as can be seen from Fig.\ref{FigHall}.\

In relatively weak  magnetic fields, $|r(H)| \ll \nu_0/\omega_H \ll 1$, the MB is negligible, quasi-particles (holes) move along the diamond-shaped orbits (see Fig.\ref{FigOrbitNet}) and hence the Hall coefficient
%
\begin{eqnarray}
R_H \approx \frac{1}{e n^{(h)}}
\label{Hallcoefficienth}
\end{eqnarray}
%
is positive (one should bear in mind that in all our expressions $e$ denotes the absolute value of electron charge).\ 

After that, an increase of magnetic field results in an increase of the MB amplitude $|r(H)|$ to become comparable to $|t(H)|$ and (still in the limit $t_0\hbar\omega_H \gg 1$) the Hall coefficient is
%
\begin{eqnarray}
R_H =0.
\label{HallcoefficientZero}
\end{eqnarray}
%
 The equality Eq.(\ref{HallcoefficientZero}) is based on the symmetry of the MB net. Indeed, changing the integral variables in Eq.(\ref{sigma3}) 
%
\begin{eqnarray}
P_x =P_x^{\prime}+ \frac{b_H^2}{b_y}(\pi/2 - \mu +\nu) \nonumber \\
P_y =P_x^{\prime}- \frac{b_H^2}{b_y}(\pi/2 +\mu +\nu)
\label{symmetry}
\end{eqnarray}
%
together with a change $P_x \rightarrow -P_x$ (or $P_y \rightarrow -P_y$) one obtains Eq.(\ref{HallcoefficientZero}).

The Hall coefficient vanishes in the whole interval of magnetic fields for which $W \gg \hbar\nu_0$, as the quasi-particle is effectively delocalized in both directions of the 2D MB net.\

With further increase of magnetic field, in the limit $|t(H)| \ll \nu_0/\omega_H \ll 1$, the Hall coefficient restores to nonvanishing value, changing its sign from  positive to negative because the quasi-particles  move along the electron-like circular orbits in this limit, i.e.
%
\begin{eqnarray}
R_H \approx - \frac{1}{e n^{(e)}}.
\label{Hallcoefficiente}
\end{eqnarray}
%

%
%

\section{Conclusion \label{SecConclusion}}

Quantum dynamics and kinetics of quasi-particles, under strong magnetic field, in the 2D net of diamond-shaped semiclassical trajectories, generated by the reconstruction of initially circular Fermi surface by the biaxial density wave,  coupled by the magnetic breakdown (MB) are considered. We present the first analytical solution of the 2D MB problem, within the semiclassical approximation, with respect to the present literature known to us. We find that the obtained quasi-particle spectrum and wave functions of Bloch type, in the momentum representation, are functions of the conserved 2D generalized momentum $\mathbf{P}=(P_x,P_y)$ if magnetic field satisfies the rational magnetic flux quantization rule analogous to the Zak/Hofstdater condition. The quasi-particle spectrum is found to be the series of alternating 2D energy bands of the width proportional to the magnetic energy $\hbar \omega_H$, and gaps between them, both in detail determined by the MB amplitude of the quasi-particle tunneling between the semiclassical trajectories.\

Using the explicit expression for the dispersion equation, we find that the diagonal magnetoconductivity $\sigma_{xx}$ and $\sigma_{yy}$ oscillate with a giant amplitude when magnetic field is changed. Period of these oscillations with respect to the inverse magnetic field is proportional to the area of the pocket in the momentum space encircled by the semiclassical quasi-particle trajectory.
We also show that the non-diagonal conductivity is equal to zero, i.e. $\sigma_{xy}=\sigma_{yx}=0$ in wide range of magnetic fields in which the width of the quasi-particle band is much larger than the energy uncertainty caused by the quasi-particle - impurity scattering, and in which the MB probabilities for a quasi-particle to move along the semiclassical trajectories and to tunnel between them are comparable. In such conditions, the quasi-particle moves freely in $x$- and $y$-direction in the 2D net as if there is effectively no magnetic field acting on it to cause the Hall effect, consequently resulting in vanishing Hall coefficient. 

In the opposite limit of negligibly narrow band-width with respect to the impurity scattering energy uncertainty, the spectrum is a set of discrete Landau levels and the magnetoconductivity tensor is the conventional one for closed trajectories, well-known from the textbooks. For relatively weak fields, when small MB tunneling probability yields the preferable motion along the closed diamond-shaped hole orbits, we have standard expression for the Hall coefficient, inversely proportional to concentration of carriers (holes), with positive sign. On the other hand, for relatively strong magnetic fields,  when strong MB tunneling probability between hole orbits reconstructs the preferable motion of carriers along the closed electron-like circular orbits, we have again the standard expression for the Hall coefficient, inversely proportional to concentration of carriers (electrons), but now with negative sign. Altogether, we have the hole-like Hall effect for weak field, switching to the electron-like one for strong fields, with wide interval of magnetic field with vanishing Hall effect between them.
Our preliminary estimations show that the analytical magnetoconductivity model suggested here may be used to explain the absence of the Hall effect and change of its sign observed in certain experiments in high-$T_c$ superconducting cuprates \cite{HallAbsence, HallChange}.

\section{Acknowledgments \label{SecAcknoledg}}

This work was supported by the Croatian Science Foundation, project IP-2016-06-2289, and by the QuantiXLie Centre of Excellence, a project cofinanced by the Croatian Government and European Union through the European Regional Development Fund - the Competitiveness and Cohesion Operational Programme (Grant KK.01.1.1.01.0004).\\

correspondence:\

*AMK: anatolykadigrobov@gmail.com\

$\dagger$DR: dradic@phy.hr

\bigskip

\begin{appendix}

\section{General formula for current of delocalized quasi-particles  
\label{AppCurrent}}

Introducing the symmetric operator of electron density at a point $\mathbf{R}$ in the coordinate space,
%
\begin{eqnarray}
\hat{n}(\mathbf{R})=\frac{1}{2}\left[\hat{\rho}\delta(\mathbf{\hat{r}}-\mathbf{R}) + \delta(\mathbf{\hat{r}}-\mathbf{R})\hat{\rho} \right],
\label{dencoperator}
\end{eqnarray}
%
one writes the local current density at the point $\mathbf{R}=(X,Y)$ in the form
%
\begin{eqnarray}
\mathbf{j}(\mathbf{R})=-2 e \mathrm{Tr}\left(\hat{\mathbf{v}}\hat{n}(\mathbf{R})  \right),
\label{currentlocal}
\end{eqnarray}
%
factor 2 coming from the quasi-particle spin.\

Writing the trace in the representation of the proper functions of Hamiltonian $\hat{H}_0$ and using Eq.(\ref{velocityME}) one finds the current density
%
\begin{eqnarray}
j_i(\mathbf{R})=-2 e^2 t_0  \sum_\kappa  E_k \frac{\partial \varepsilon_\kappa}{\partial P_k} \frac{\partial \varepsilon_\kappa}{\partial P_i}  \frac{\partial f_0}{\partial \varepsilon_\kappa}\varphi_\kappa^\star(\mathbf{R})\varphi_\kappa(\mathbf{R})
\label{currenR}
\end{eqnarray}
%
where $\mathbf{E}$ is electric field and $i \in \{ x,y \}$.
Here  $\varphi_{\kappa}$ is the electron proper Bloch functions
%
\begin{eqnarray}
\varphi_{n,\mathbf{P}}(\mathbf{R})=e^{i\mathbf{P}\mathbf{R}/\hbar}u_{n,\mathbf{P} }(\mathbf{R})
\label{Blochfunctionproper}
\end{eqnarray}
%
in the coordinate representation, $\kappa=(n,\mathbf{P})$, i.e. the band index and conserved generalized momentum respectively.
The total current flowing along the plate of the length $L_x$ and the width $L_y$  is 
%
\begin{eqnarray}
\mathbf{I}=\int^{L_x/2}_{-L_x/2} dX\int_{-L_y/2}^{L_y/2}dY\; \mathbf{j}(\mathbf{R}).
\label{totalcurrenR}
\end{eqnarray}
%
Using the normalization condition for the periodic factors in Bloch functions, one finds
%
\begin{eqnarray}
\int^{L_x/2}_{-L_x/2} dX\int_{-L_y/2}^{L_y/2}dY u_{n,\mathbf{P} }^{\star}(\mathbf{R}) u_{n,\mathbf{P} }(\mathbf{R})=L_X L_Y
\label{normalizationU}
\end{eqnarray}
%
and, hence, the current density averaged over the sample, $\mathbf{J} =\mathbf{I}/L_xL_y$, is
%
\begin{eqnarray}
J_i = -2 e^2 t_0 \sum_{n} E_k \int \frac{d \mathbf{P}}{(2 \pi \hbar)^2} \frac{\partial
 \varepsilon_n}{\partial P_i}  
 \frac{\partial \varepsilon_n}{\partial P_k} 
\frac{\partial f_0}{\partial \varepsilon}\Big|_{\varepsilon=\varepsilon_n(\mathbf{P})}.\nonumber\\
 \label{AppAverageCurr}
\end{eqnarray}
%
 
\section{Calculation of current in terms of dispersion function \label{AppDcurrent}}

Solution of the dispersion equation Eq.(\ref{dispersionequation}) appearing in  Eq.(\ref{velocityD}) is $\varepsilon=\varepsilon_n(\mathbf{ P})$, thus the dependence of  $v_i$  on $\mathbf{ P}$ is $v_i(\mathbf{P}) =v_i(\varepsilon_n(\mathbf{ P}),\mathbf{ P})$. The latter dependence allows to rewrite  Eq.(\ref{sigma}) in the form
%
\begin{eqnarray}
\sigma_{ik} &=& -2 e^2 t_0 \nonumber\\
&\times & \int \frac{d \mathbf{P}}{(2 \pi \hbar)^2} \int d \varepsilon \frac{\partial f_0}{\partial \varepsilon} v_i(\varepsilon,\mathbf{P} ) v_k(\varepsilon,\mathbf{P} )\sum_n 
\delta \left[\varepsilon -\varepsilon_n(\mathbf{ P}) \right]. \nonumber\\
\label{AppSigma1}
\end{eqnarray}
%
Using the known mathematical identity
%
\begin{eqnarray}
\delta[f(x)]=\sum_k \frac{\delta[x-x_k]}{f^\prime (x)},
\label{deltaidentity}
\end{eqnarray}
%
where $x_k$ are zeros of function $f(x)$, and Eq.(\ref{velocityD}), one finds Eq.(\ref{sigma1}) from the main text.

\end{appendix}

\begin{thebibliography}{99}
\bibitem{Comment} Topological  reconstruction of closed FSs into a 1D chain of FSs was suggested and its stability investigated in our recent papers \cite{DWPRB1,DWPRB2,DWPRB3}.
\bibitem{comment} If the magnetic field corresponds to the case that the Hamiltonian commutes with the operator of translation on $n b_x$ and $b_y$ (here $n \geq 1$),
the term $2 \pi l $ in the  right-hand side of Eq.(\ref{ZakCond1}) is changed to $2 \pi l/n $ and hence the flux ratio is $\Phi_0/\Phi = p/q$
with $p$ and $q$ integers for all rational magnetic fields.
\end{thebibliography}
\end{document}